# Magnetic and dielectric properties of multiferroic $Eu_{0.5}Ba_{0.25}Sr_{0.25}TiO_3$ ceramics

Veronica Goian,[a] Stanislav Kamba,[a] Přemysl Vaněk,[a] Maxim Savinov,[a] Christelle Kadlec,[a] Jan Prokleška[b]

[a]*Institute of Physics, Academy of Sciences of the Czech Republic, Na Slovance 2, 18221 Prague 8, Czech Republic*

[b]*Faculty of Mathematics and Physics, Department of Condensed Matter Physics, Charles University, Ke Karlovu 5, 121 16, Prague 2, Czech Republic*

Abstract:

Dielectric and magnetic properties of $Eu_{0.5}Ba_{0.25}Sr_{0.25}TiO_3$ are investigated between 10 K and 300 K in the frequency range from 10 Hz to 100 THz. A peak in permittivity revealed near 130 K and observed ferroelectric hysteresis loops prove the ferroelectric order below this temperature. The peak in permittivity is given mainly by softening of the lowest frequency polar phonon (soft mode revealed in THz and IR spectra) that demonstrates displacive character of the phase transition. Room-temperature X-ray diffraction analysis reveals cubic structure, but the IR reflectivity spectra give evidence of a lower crystal structure, presumably tetragonal *I4/mcm* with tilted oxygen octahedra as it has been observed in $EuTiO_3$. The magnetic measurements show that the antiferromagnetic order occurs below 1.8 K. $Eu_{0.5}Ba_{0.25}Sr_{0.25}TiO_3$ has three times lower coercive field than $Eu_{0.5}Ba_{0.5}TiO_3$, therefore we propose this system for measurements of electric dipole moment of electron.

## 1. Introduction

The materials which exhibit two or more ferroic properties (anti/ferroelectricity, magnetic order and/or strain) are called multiferroics. The possibilities of coupling between dipolar and magnetic orders i.e. magnetoelectric coupling are appealing due to their potential applications in non-volatile memories, high sensitive ac magnetic field sensors, etc [1,2,3]. However, the multiferroics with such coupling are rare in nature, the coupling is usually very weak and it occurs mainly at temperatures deeply below room temperature [4]. Therefore, there is an intensive search for new multiferroics with high critical temperatures and with strong magnetoelectric coupling.

On the other hand, the low-temperature multiferroics were very recently suggested for the tests of cosmological models [5] or even elementary particle theories [6,7]. The standard model of particles assumes among others charge and parity symmetry violation and one of the consequences of this violation is the prediction that each elementary particle should have a small electric dipole moment (EDM). For example according to the Standard model electrons should have EDM of order $\approx 10^{-40}$ $e$.cm ($e$ marks the charge of the electron) [8]. However, there are experimental evidences that the charge-parity symmetry violation is much larger than the Standard model predicts [9,10]. There are many theories going beyond the Standard model now, but unfortunately, the tests of such theories are very difficult. One of the possibilities is the determination of EDM, whose value depends on the type of theory. The physicists have tried to measure EDM already for more than forty years [11,12], but the highest sensitivity achieved up to now is only $10^{-26}$ $e$.cm [13]. Recently, it was proposed [6], that the use of multiferroic $Eu_{0.5}Ba_{0.5}TiO_3$ can enhance the sensitivity up to $10^{-27}$ $e$.cm. Internal electric fields in the order of 10 MV/cm present in this material can be easily switched by an external electric field of ~10 kV/cm even at low temperatures. Large internal electric fields should affect the spins of 7 electrons in the $f$ shell of Eu and if the switching of the electric field will influence EDM, linear magnetoelectric effect should be observed in the paramagnetic phase of $Eu_{0.5}Ba_{0.5}TiO_3$, although it is forbidden by crystal symmetry. Details of such experiment go beyond the scope of this manuscript, but they can be found in Ref. 7.

$Eu_{0.5}Ba_{0.5}TiO_3$ is a solid solution of $EuTiO_3$ [14] and $BaTiO_3$ [15]. $EuTiO_3$ is an antiferromagnetic material with G-type ordering below 5.3 K. It is a quantum paraelectric, which exhibits a large dielectric constant at low temperatures and its dielectric response is strongly affected by a magnetic order [14] or by a magnetic field [16]. In contrast, $BaTiO_3$ is a prototype ferroelectric with a large room temperature polarization [15]. By substituting partially Ba on the Eu site of $EuTiO_3$, the magnetic ordering will be suppressed through the dilution and simultaneously $Eu_{1-x}Ba_xTiO_3$ will become ferroelectric through the expansion of the lattice constant. The aim is to find a multiferroic with a low Néel temperature, but with a high concentration of magnetic ions. The system should have simultaneously large polarization (i.e. high internal electric field), which is switchable at low temperatures. $Eu_{0.5}Ba_{0.5}TiO_3$ seems to fulfil such conditions. It is ferroelectric below 215 K with *Amm2* symmetry and it becomes antiferromagnetic below 1.9 K [17]. The complete ferroelectric phase diagram of $Eu_{1-x}Ba_xTiO_3$ has been recently published [18]. Unfortunately, $Eu_{0.5}Ba_{0.5}TiO_3$ exhibits large coercive field of 10 kV/cm. The sample heats during the polarization switching, which reduces the sensitivity of EDM measurements down to $10^{-25}$ $e$.cm [19]. The design of a material with a lower ferroelectric transition temperature would permit to manipulate with smaller coercive fields. This can be achieved for example by replacing of Ba by Sr in $Eu_{0.5}Ba_{0.5}TiO_3$. In such case the magnetic properties should not be affected.

The aim of this manuscript is to describe the sintering of $Eu_{0.5}Ba_{0.25}Sr_{0.25}TiO_3$ ceramics and characterization of their dielectric and magnetic properties. We will show that the coercive field of $Eu_{0.5}Ba_{0.25}Sr_{0.25}TiO_3$ is three times smaller than in $Eu_{0.5}Ba_{0.5}TiO_3$ and that the AFM order occurs at the same temperature 1.9 K like in $Eu_{0.5}Ba_{0.5}TiO_3$. Due to these properties $Eu_{0.5}Ba_{0.25}Sr_{0.25}TiO_3$ is appear to be more suitable for EDM measurements than $Eu_{0.5}Ba_{0.5}TiO_3$.

## 2. Experimental techniques:

$Eu_{0.5}Ba_{0.25}Sr_{0.25}TiO_3$ was synthesized by solid-state reaction using mechanochemical activation before calcination. $Eu_2O_3$, $TiO_2$ (anatase), $BaTiO_3$ and $SrTiO_3$ powders (all from Sigma Aldrich) were mixed in stoichiometric ratio, then milled intensively in a Fritsch Pulverisette 7 planetary ball micromill for 120 min in a dry environment followed by 20 min milling in suspension with *n*-heptane. $ZrO_2$ grinding bowls (25 ml) and balls (12 mm diameter, acceleration 14 g or 137 m/s$^2$) were used. The suspension was dried under an infrared lamp and the dried powder was pressed in a uniaxial press (330 MPa, 3 min) into 13-mm-diameter pellets. The pellets were calcined in a pure $H_2$ atmosphere at 1200°C for 24 h (to reduce $Eu^{3+}$ to $Eu^{2+}$), then milled and pressed by the same procedure as above and sintered at 1300°C for 24 h in Ar + 10% $H_2$ atmosphere.

Low temperature IR reflectivity spectra were obtained using a Fourier transform IR spectrometer Bruker IFS 113v in the frequency range 30–650 cm$^{-1}$, room temperature spectra were taken up to 3000 cm$^{-1}$. He-cooled Si bolometer operating at 1.6 K was used as a detector below 650 cm$^{-1}$ for measurements at low temperatures. For higher frequencies a pyroelectric deuterated triglycine sulfate detector was used. Time-domain THz transmission spectroscopy measurements were performed in the range of 100 GHz to 1.5 THz. Linearly polarized THz probing pulses were generated using a Ti:sapphire femtosecond laser whose pulses illuminated an interdigitated photoconducting GaAs switch. The THz signal was detected using electro-optic sampling with a 1 mm thick [110] ZnTe crystal. An Optistat CF Oxford Instruments continuous flow helium cryostat was used in both THz and IR experiments for cooling the samples down to 10 K.

Low-frequency (10 Hz – 1 MHz) dielectric measurements were performed between 10 and 300 K using NOVOCONTROL Alpha-A High Performance Frequency Analyzer. The ferroelectric hysteresis loops were measured at frequencies of 1-50 Hz and temperatures between 10 and 150 K. At higher temperatures the ceramics were too conducting, so that only dielectric lossy loops were detected.

Magnetic susceptibility and magnetization were obtained using a Quantum Design PPMS9 and a $He^3$ insert equipped with a home-made induction coil that enabled measurements of ac magnetic susceptibility, $\chi$, from 0.5 to 300 K at 0.1 T.

## 3. Results and discussion

In order to identify the ferroelectric phase transition in $Eu_{0.5}Ba_{0.25}Sr_{0.25}TiO_3$ we measured the complex dielectric permittivity and ferroelectric hysteresis loops (Figure 1). The low temperature permittivity, obtained between 10 Hz and 0.6 MHz, shows a maximum near 130 K. This feature together with the observed ferroelectric loops gives evidence of a ferroelectric phase transition at $T_C=130$ K. The permittivity increases rapidly on heating above $T_C$ as a consequence of the conductivity and related Maxwell-Wagner polarization. The peak in the real part of the permittivity at 10 Hz is partially hidden by the conductivity near $T_C$, but above 80 Hz we clearly see the intrinsic dielectric permittivity near $T_C$. Such a behaviour is also observed in dielectric loss. Below 150 K, we have an intrinsic dielectric relaxation (tan$\delta$ <0.1) stemming probably from the domain wall motion, but at higher temperatures the loss increases due to the conductivity, and tan$\delta$ exceeds unity at temperatures higher than 150 K (see Figure 1b). In the inset of Figure 1 two hysteresis loops are plotted at 11 and 120 K. As expected, the coercive field increases on cooling up to a value of 3.3 kV/cm and the saturated polarization reaches value of 4 $\mu C/cm^{-2}$ at 11 K.

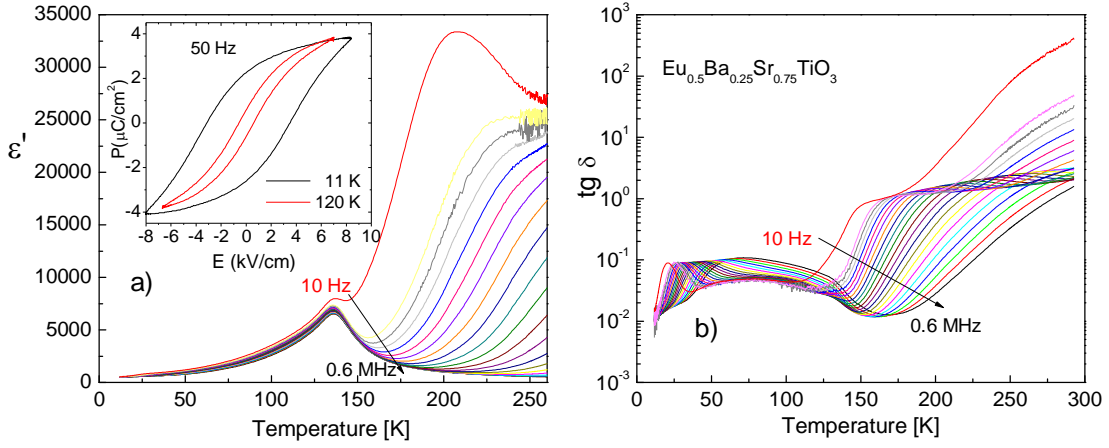

Figure 1. a) Temperature dependence of the real part of permittivity in the range of 10 Hz to 0.6 MHz. Inset shows ferroelectric hysteresis loops at 120 and 11 K. b) Temperature dependence of dielectric loss at various frequencies.

In Figure 2a the IR reflectivity spectra of $Eu_{0.5}Ba_{0.25}Sr_{0.25}TiO_3$ taken below room temperature are plotted. No dramatic change with temperature is seen on cooling: only a small increase of the intensity due to the decrease of phonon damping occurs. Note also the remarkable increase of the THz reflectivity on cooling down to 120 K and its much lower value at 10 K. However, the spectra exhibit more than three reflectivity bands specific for the cubic $Pm\bar{3}m$ perovskite structure, i.e. the symmetry is probably lower than cubic. This behavior might be due to an antiferrodistortive orthorhombic structure with *I4/mcm* space group, which is known from pure $EuTiO_3$ [20]. Nevertheless, the first preliminary XRD measurements suggest that the structure is cubic at room temperature with a lattice parameter a=3.937(6) Å. A second possible explanation of multiband reflectivity structure is multi-mode behavior given by three cations in perovskite A sites.[21] Near-normal IR reflectivity can be expressed as

$$R(\omega) = \left| \frac{\sqrt{\varepsilon^*(\omega)} - 1}{\sqrt{\varepsilon^*(\omega)} + 1} \right|^2$$

where the complex dielectric function $\varepsilon^*(\omega)$ can be expressed using a generalized oscillator model with factorized form of the complex permittivity:

$$\varepsilon^*(\omega) = \varepsilon'(\omega) - i\varepsilon''(\omega) = \varepsilon_\infty \prod_j \frac{\omega_{LOj}^2 - \omega^2 + i\omega\gamma_{LOj}}{\omega_{TOj}^2 - \omega^2 + i\omega\gamma_{TOj}}$$

where $\omega_{TOj}$ and $\omega_{LOj}$ are transverse and longitudinal frequencies of the j-th polar phonon, respectively, $\gamma_{TOj}$ and $\gamma_{LOj}$ are their damping constants, and $\varepsilon_\infty$ denotes the high frequency permittivity resulting from electronic absorption processes. The IR reflectivity spectra are less accurate below 50 cm$^{-1}$, therefore more reliable data were obtained from THz spectra. For that reason, IR reflectivity spectra were fitted simultaneously with the THz spectra.

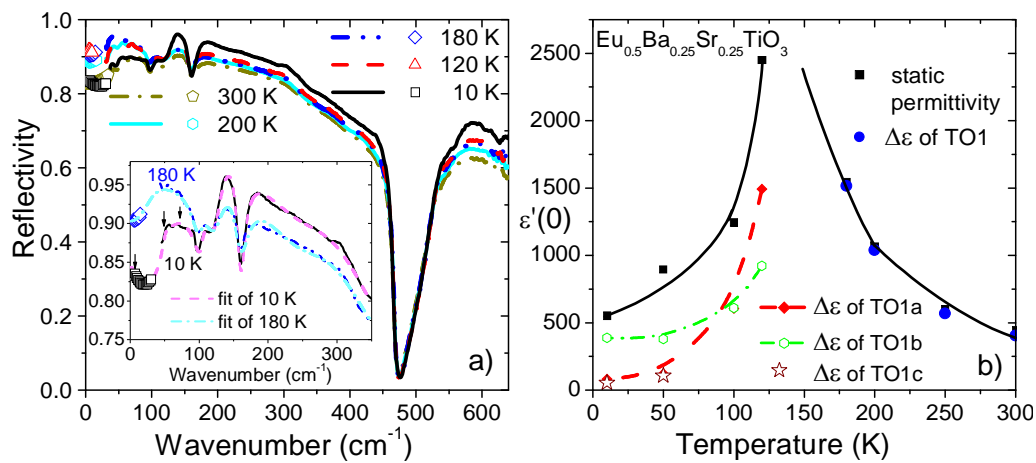

Figure 2. a) Infrared reflectivity spectra of Eu$_{0.5}$Ba$_{0.25}$Sr$_{0.25}$TiO$_3$ ceramics at selected temperatures. The inset shows two low frequency reflectivity spectra at 10 and 180 K together with their fits below 350 cm$^{-1}$. The black arrows mark the three-components of the soft mode in IR reflectivity spectra at 10 K. b) Temperature dependence of the static permittivity obtained from the sum of all phonon contributions. Dielectric contributions of the soft mode and its components below T$_C$ are shown in the same figure.

The behavior of the transverse polar phonon frequencies with temperature is plotted in Figure 3a. The soft mode exhibits a minimum near 130 K, i.e at T$_C$. The apparent gap is seen in the data between 180 K and 130 K due to a lack of THz data. The reason for this is that the sample becomes opaque in the THz region in this temperature range. The modes above 100 cm$^{-1}$ do not exhibit any dramatic changes in the IR spectra with the temperature. The most important part for the dynamics of the phase transition occurs below 100 cm$^{-1}$. One can see a splitting of the TO1 soft mode into two components below 130 K which harden on cooling. The lower frequency component is resolved in the THz spectra (see Figure 2b). A third component of the soft mode starts to appear below 50 K. The existence of the soft mode frequency which exhibits a minimum near T$_C$, demonstrates the displacive character of the ferroelectric phase transition.

The activation of two-components of the soft mode shows a tetragonal ferroelectric structure between 50 and 130 K. Appearance of the third component near 50 cm$^{-1}$ demonstrates an orthorhombic structure below 50 K. The dielectric strengths of the soft mode and its components below $T_C$ are shown in inset of Figure 1. The static permittivity $\varepsilon$'(0) is given by the sum of all phonon contributions and the electronic contributions (see inset of figure 1)

$$\varepsilon(0) = \sum_{j=1}^{n} \Delta\varepsilon_j + \varepsilon_\infty$$

The static permittivity exhibits a maximum near 130 K as consequence of the ferroelectric phase transition. The dispersion in permittivity is observed in Figure 1. The value of static permittivity $\varepsilon$'(0) exhibits a drastic change with the temperature due to the change of the soft mode frequency with temperature. At $T_C$ the permittivity decreases from 8000 at 10 Hz down to 6400 at 0.6 MHz. The static permittivity from phonon contributions is about 3000 at $T_C$. It means that some dielectric relaxation lying below the phonon frequencies contributes to the permittivity. A similar behaviour was also observed in $Eu_{0.5}Ba_{0.5}TiO_3$ [17]. It was explained by the presence of some $Eu^{3+}$ cations or oxygen vacancies in the lattice. Another possible explanation of the dielectric relaxation is the presence of some dynamical disorder in the lattice close to $T_C$. In that case the phase transition would manifest a crossover between displacive and order-disorder character. Above $T_C$ the presence of defects is responsible for the huge conductivity which induces an increase of the permittivity observed in Figure 1.

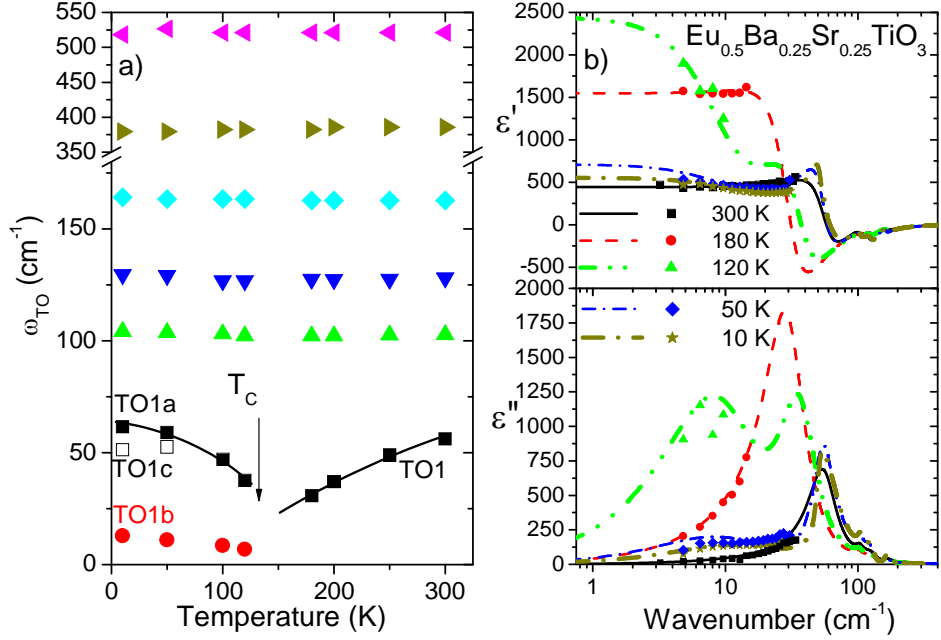

Figure 3. a) Temperature dependence of the transverse optical phonon frequencies in $Eu_{0.5}Ba_{0.25}Sr_{0.25}TiO_3$ ceramics; b) Real and imaginary parts of the complex permittivity spectra for selected temperatures. The phonon frequencies and $\varepsilon^*(\omega)$ spectra were obtained from the fit of IR reflectivity spectra.

In Figure 3b, we show the real and imaginary components of the complex dielectric spectra below room temperature. Symbols are experimental THz data and lines are the results of the THz and IR fits. The peaks in imaginary part of the permittivity correspond to the eigenfrequencies of the transverse optical phonons.

From previous measurements of $Eu_{0.5}Ba_{0.5}TiO_3$, we know that antiferromagnetic order occurs around 1.9 K [6]. Therefore, the magnetic measurements of $Eu_{0.5}Ba_{0.25}Sr_{0.25}TiO_3$ were performed at very low temperatures. The ac susceptibility measurement exhibits a peak around 1.8 K (see Figure 4) which disappears when a static magnetic field is applied. It seems that the presence of Sr ions does not influence too much the magnetic order, since the Néel temperature in $Eu_{0.5}Ba_{0.25}Sr_{0.25}TiO_3$ is only by 0.1 K lower than in $Eu_{0.5}Ba_{0.5}TiO_3$. The reason is the same concentration of magnetically active Eu ions in both materials.

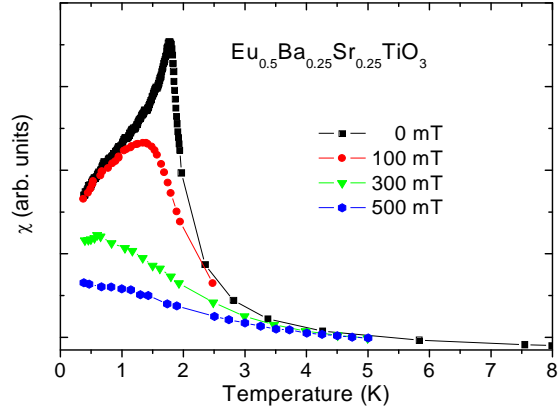

Figure 4. Temperature dependence of the magnetic susceptibility for different applied magnetic fields.

## 4. Conclusions

Low frequency dielectric measurements demonstrated that $Eu_{0.5}Ba_{0.25}Sr_{0.25}TiO_3$ ceramics is ferroelectric below 130 K. At 10 K, the ferroelectric hysteresis loops were clearly seen with a coercive field of 3.3 kV/cm and spontaneous polarization of 4 μC/cm$^{-2}$. The ferroelectric phase transition is driven by the ferroelectric soft mode observed in the THz complex permittivity and IR reflectivity spectra. Additional dielectric relaxation below phonon frequencies is observed which could be responsible for the higher values of the permittivity in the kHz range. It means that the phase transition is mainly displacive but also some order-disorder component exists. The infrared spectra show that paraelectric phase has a symmetry lower than cubic. Most probably, it is tetragonal *I4/mcm* as in $EuTiO_3$. The magnetic Néel temperature $T_N$ = 1.8 K is only slightly affected by the presence of Sr ions, as $Eu_{0.5}Ba_{0.5}TiO_3$ has $T_N$ just by 0.1 K higher. At low temperatures the magnitude of the ferroelectric coercive field is three times lower than in $Eu_{0.5}Ba_{0.5}TiO_3$, therefore one can expect that $Eu_{0.5}Ba_{0.25}Sr_{0.25}TiO_3$ is more suitable for measurements of the electric dipole moment of electron than $Eu_{0.5}Ba_{0.5}TiO_3$.

## Acknowledgements

This work was supported by the Czech Science Foundation (Project No. P204/12/1163) and MŠMT (COST MP0904 project LD12026).